\documentclass[12pt]{iopart}

\usepackage{graphicx}
\usepackage{dcolumn}
\usepackage{bm}

\begin{document}
\title[Terahertz conductivity of localized photoinduced carriers in Mott insulator YTiO$_{3}$]{Terahertz conductivity of localized photoinduced carriers in Mott insulator YTiO$_{3}$ at low excitation density, contrasted with metallic nature in band semiconductor Si}

\author{J. Kitagawa$^1$, Y. Kadoya$^1$, M. Tsubota$^2$, F. Iga$^1$ and T. Takabatake$^1$}

\address{$^1$ Department of Quantum Matter, ADSM, Hiroshima University, 1-3-1 Kagamiyama, Higashi-Hiroshima 739-8530, Japan}
\address{$^2$ Synchrotron Radiation Research Unit, JAEA, Hyogo 679-5148, Japan}
\ead{jkita@hiroshima-u.ac.jp}

\begin{abstract}
We performed optical-pump terahertz-probe measurements of a Mott insulator YTiO$_{3}$ and a band semiconductor Si using a laser diode (1.47 eV) and a femtosecond pulse laser (1.55 eV). Both samples possess long energy-relaxation times (1.5 ms for YTiO$_{3}$ and 15 $\mu$s for Si); therefore, it is possible to extract terahertz complex conductivities of photoinduced carriers under equilibrium. We observed highly contrasting behavior - Drude conductivity in Si and localized conductivity possibly obeying the Jonscher law in YTiO$_{3}$. The carrier number at the highest carrier-concentration layer in YTiO$_{3}$ is estimated to be 0.015 per Ti site. Anisotropic conductivity of YTiO$_{3}$ is determined. Our study indicates that localized carriers might play an important role in the incipient formation of photoinduced metallic phases in Mott insulators. In addition, this study shows that the transfer-matrix method is effective for extracting an optical constant of a sample with a spatially inhomogeneous carrier distribution.
\end{abstract}

\submitto{\JPCM}
\maketitle

\clearpage

\section{Introduction}
Recent discoveries of photoinduced metallic phases in several Mott insulators\cite{Miyano:PRL,Cavalleri:PRL,Iwai:PRL,Tajima:JPSJ,Chollet:Science,Perfetti:PRL} made us consider the strongly correlated electron physics from a new point of view.
However, the optical properties of photoinduced carriers in Mott insulators are not well understood even at low excitation densities.
Understanding these optical properties is a prerequisite to understanding the incipient creation of metallic phases.
Drude response by itinerant carriers is observed in the case of band semiconductors with low excited-carrier density (10$^{14}$$\sim$10$^{16}$ cm$^{-3}$)\cite{Ralph:PRB,Beard:PRB,Shan:PRL,Dakovski:JOSAB}.
The comparison between the optical properties of photoinduced carriers at low excitation densities in Mott insulators and those in band semiconductors would be important in gaining deeper insight into strongly correlated electron physics.

The detailed nature of various carrier conductions, exhibiting Drude or hopping conduction, can be well characterized in the terahertz (THz) regime\cite{Exter:APL,Katzenellenbogen:APL,Jeon:PRL,Harimochi:PRB}.
THz time-domain spectroscopy (THz-TDS) is a powerful tool for analysing terahertz conductivity $\tilde{\sigma}(\omega)$ ($=\sigma_{1}(\omega)+i\sigma_{2}(\omega)$).
The remarkable advantage of THz-TDS is its simultaneous determination of both the real and imaginary parts of $\tilde{\sigma}(\omega)$, without using the Kramers-Kronig transformation\cite{Gruner:MS}.
The coherent nature of the THz pulse is also utilized to investigate photoinduced $\tilde{\sigma}(\omega)$ by, for instance, optical-pump THz-probe (OPTP) studies.
Recent progress in THz technologies\cite{Mittleman,Sakai} and the methodology of analysis\cite{Kindt:JCP,Nemec:JCP,Nienhuys:PRB} in OPTP experiments enable evaluation of transient $\tilde{\sigma}(\omega)$ in many substances, such as semiconductors, high-T$_{c}$ superconductors, liquids and organic materials\cite{Beard:PRB,Averitt:JOSAB,Huber:Nature,Averitt:PRL,Beard:JAP,Knoesel:PRL,Hegmann:PRL,Turner:JPCB,Demsar:PRL,Kaindl:Nature,Hendry:PRB2,Kampfrath:PRL}.

Because the OPTP method essentially detects non-equilibrium processes, such as surface recombination and carrier diffusion, the time dependence of a complicated spatial carrier distribution must be considered\cite{Beard:PRB,Gallant:PRB}.
This difficulty is avoided by using thin-film samples\cite{Averitt:JOSAB,Huber:Nature,Averitt:PRL,Beard:JAP,Demsar:PRL,Kampfrath:PRL} where the optical pump pulse penetrates, and by analysing the photoinduced phase as a layer with a homogeneous $\tilde{\sigma}(\omega)$\cite{Beard:PRB,Huber:Nature,Beard:JAP,Demsar:PRL,Schall:OL}.
Furthermore the extraction of $\tilde{\sigma}(\omega)$, varying quickly compared with the pulse width of THz probe pulse, only seems possible within some restricted conditions\cite{Kindt:JCP,Nemec:JCP,Nienhuys:PRB}.
Therefore, to analyse photoinduced $\tilde{\sigma}(\omega)$ for a wide variety of Mott insulators, free from the restrictions in sample preparation and analysis, we initially examined a nearly equilibrium state of photoexcited bulk material with a longer energy-relaxation time, $\tau$.
The photoexcitation of a material with longer $\tau$ creates a quasi-equilibrium state averaging over various non-equilibrium processes, making it easy to obtain the optical constants of the highest carrier-concentration region in the material.

Our analysis also required extracting $\tilde{\sigma}(\omega)$ of materials with inhomogeneous carrier distributions in a more rigorous manner.
One good candidate for accomplishing this is the transfer-matrix method, which expresses inhomogeneous carrier distribution by a multi-layer system.
This method has been briefly commented on in the literature\cite{Beard:PRB}.
The promise that the transfer-matrix method can incorporate inhomogeneity is seen in the analyses of reflectivity in optical pump-probe studies\cite{Ma:JAP,Okamoto:PRB}.
Although it is a versatile method, its effectiveness has not been thoroughly discussed, especially in THz-TDS studies.

In this study, we found that $\tau$ of a Mott insulator YTiO$_{3}$ with Mott gap of approximately 1 eV\cite{Taguchi:PRB} is 1.5 ms at 1.47 eV photoexcitation.
We characterized the photoinduced $\tilde{\sigma}(\omega)$ by comparing it with that of a band semiconductor Si with a bandgap of 1.1 eV\cite{Sze}.
We also present a more detailed discussion of the transfer-matrix method.

\section{Experimental method}

Single-crystalline samples of YTiO$_{3}$, with the orthorhombic perovskite GdFeO$_{3}$-type structure, were grown by the floating zone method\cite{Tsubota:JPSJ}.
The Si sample was commercial high-resistivity Si.

Photoconductivity was measured to assess $\tau$.
The light emitted from a multimode continuous wave (CW) laser diode (LD) with photon energy of 1.47 eV was modulated and used for illuminating the sample under an electric field of about 0.3 kV/cm.
The photocurrent $I_{ph}$ flowing through a 100 $\Omega$ resistance, connected in series with the sample, was lock-in detected.

OPTP experiment was performed by a transmission THz-TDS system described in detail elsewhere\cite{Kitagawa:JPSJ,Kitagawa:JMMM}.
The thicknesses of platelet samples are 420 $\mu$m for YTiO$_{3}$ and 512 $\mu$m for Si.
The optical pulses were generated by a mode-locked Ti-sapphire laser with a repetition rate of 76 MHz and central wavelength of 800 nm.
Both the THz emitter and detector were low-temperature grown GaAs photoconductive antennas.
Si lenses were attached to the antennas to enhance the emission power and collection efficiency of THz pulses.
The THz spectral range in this experiment was between 0.5 and 8 meV.
The THz-wave-emission sides of the samples were photoexcited by the multi-mode CW LD with an incident angle of 45$^{\circ}$.
The pump-beam power was 0.8 W for Si and 1.2 W for YTiO$_{3}$.
For YTiO$_{3}$, the polarization of THz electric field $E_{THz}$ is parallel to the {\it b}-axis.
The beam diameter of the LD light was about 8 mm and larger than that of the THz probe pulse, which is energy dependent (e.g. 2 mm at 2 meV and 1 mm at 4 meV).
The fluence rate was 1.6 W/cm$^{2}$ for Si and 2.4 W/cm$^{2}$ for YTiO$_{3}$, respectively.
The temperature rise resulting from the thermalisation by photoexcitation\cite{thermal} is estimated not to exceed 1 K.

OPTP measurements were also performed with the 1.55 eV optical-pump pulses split from the Ti-sapphire laser to investigate the nature of conduction of carriers induced by light with photon energy larger than that of the CW LD.
This experiment studied the anisotropy of photoinduced $\tilde{\sigma}(\omega)$ in YTiO$_{3}$.
The optical-pump pulse power was 230 mW.
The optical-pump pulse beam diameter was about 2 mm (fluence: 96 nJ/cm$^{2}$) - slightly smaller than that of the THz probe pulse below 2 meV.
The $E_{THz}$ was applied along either the $b$- or $c$-axis, maintaining the polarization of the optical-pump pulse electric field $E_{1.55eV}$ parallel to the $b$- or $c$-axis.

All measurements were performed at room temperature.

\section{Results and Discussion}

Figure 1 shows the modulation-frequency dependence of $I_{ph}$.
Increasing the modulation frequency causes $I_{ph}$ to decrease, according to $\frac{B}{(1+(\omega\tau)^2)}$, where $B$ is the proportional coefficient.
The obtained $B$ and $\tau$ are listed in the figure.
Longer $\tau$ ($\sim$ 15 $\mu$s for Si and $\sim$ 1.5 ms for YTiO$_{3}$) indicates that the samples are in quasi-equilibrium photoinduced state during the THz-TDS measurements.
$\tau$ of YTiO$_{3}$ is almost excitation-intensity independent; this implies that a thermal effect does not dominate the relaxation process.
Although the direction of $I_{ph}$ and of electric field of the CW LD pump are not identified accurately in YTiO$_{3}$, a huge anisotropic $\tau$ that depends on the direction of $I_{ph}$ and the polarization of the excitation light is not anticipated.
The OPTP results shown later support this (see Fig.\ 7).

The temporal evolution of $E_{THz}$ transmitted through Si and YTiO$_{3}$ are shown in Fig.\ 2(a) and 2(b), respectively, with and without LD excitations (1.47 eV).
Photoexcitation attenuates both THz waves implying THz wave absorption is by the photoinduced carriers.
THz energy dependence of transmission $T(\omega)$ and phase shift $\Delta\phi(\omega)$ are obtained by Fourier transformation of THz waves, as shown in Fig.\ 3(a) and 3(b).
They are calculated with the equations $T(\omega)=\frac{E(\omega)}{E_{ref}(\omega)}$ and $\Delta\phi(\omega)=\phi(\omega)-\phi_{ref}(\omega)$, where $E(\omega)(E_{ref}(\omega))$ and $\phi(\omega)(\phi_{ref}(\omega))$ are the Fourier transformed amplitude and phase, with and without excitation, respectively.
THz wave absorption by photoinduced carriers is responsible for $T(\omega)$ decreasing below 1 for both samples, but the two exhibit different energy dependencies.
As the THz energy increases, $T(\omega)$ of Si approaches 1, while that of YTiO$_{3}$ gradually decreases.
The opposite sign of $\Delta\phi(\omega)$ of the two samples strongly indicates different fundamental conduction mechanisms of the photoinduced carriers.
Negative (positive) $\Delta\phi$ roughly means the refractive index is reduced (increased), compared with an unexcited state, which influences the negative (positive) real part of the dielectric constant of photoinduced carriers.
As explained below, these results indicate a metallic nature below a plasma frequency in Si, and a localized nature, such as hopping carriers, in YTiO$_{3}$.

Before showing the photoinduced $\tilde{\sigma}(\omega)$ of Si and YTiO$_{3}$, we mention the detailed procedure of the transfer-matrix method.
A spatially inhomogeneous distribution of photoinduced carriers is initially regarded as exponentially decaying.
Subsequently a non-exponentially decaying distribution is introduced.
The photoinduced phase with an exponentially decaying carrier distribution is divided into many thin slabs (see Fig.\ 4).
Each slab is supposed to have a uniform complex refractive index $\tilde{n}_{j}$ whose value is set to reproduce the exponential decay of the photoinduced carrier concentration.
The transfer-matrix of each slab is described by
\begin{equation}
M_{j}=\left(\begin{array}{cc}
        \cos \delta&i\sin \delta/\tilde{n}_{j}\\
        i\tilde{n}_{j}\sin \delta&\cos \delta
        \end{array}\right),
\label{equ:monomatrix}
\end{equation}
where $j$ is the number index of the slab, $\delta=\frac{2\pi}{\lambda}\tilde{n}_{j}d$, $\lambda$ is the incident THz wavelength in vacuum, and $d$ is the slab thickness\cite{Born,Harimochi:JJAP}.
The $\tilde{n}_{j}$ of each slab is calculated from the complex dielectric constant $\tilde{\epsilon}$ using,
\begin{equation}
\tilde{n}_{j}^{2}=\tilde{\epsilon}_{no}+\tilde{\epsilon}_{sur}\exp\left(-\frac{z_{j}}{d_{p}}\right),
\label{equ:epsilon}
\end{equation}
where $\tilde{\epsilon}_{no}$ is $\tilde{\epsilon}$ without excitation, $\tilde{\epsilon}_{sur}$, the parameter to be optimized in this analysis, is $\tilde{\epsilon}$ resulting from carriers at the photoexcited surface of the sample, $z_{j}$ ($=j\times d$) is the depth from the photoexcited surface into the sample along the THz wave propagation, and $d_{p}$ is the optical penetration depth. 
For Si, frequency independent $\tilde{\epsilon}_{no}$\cite{Exter:PRB} of 11.7 is used.
For YTiO$_{3}$, $\tilde{\epsilon}_{no}$ is experimentally determined by the THz-TDS measurement and is weakly energy dependent (e.g. 16.5+0.4$i$ at 2 meV and 17+0.4$i$ at 4 meV).
$d_{p}$ of Si at 1.47 eV is determined to be 8.4 $\mu$m using the absorption coefficient from an optical data handbook\cite{Palik}.
That of YTiO$_{3}$ at 1.47 eV was calculated to be 0.22 $\mu$m from the reported reflectivity spectra\cite{Okimoto:PRB} (0.05-40 eV) combined with the Kramers-Kronig transformation.
Then the total matrix $M_{t}$ is described as
\begin{equation}
M_{t}=\prod_{j=k}^{0}M_{j},
\label{equ:totalmatrix}
\end{equation}
where $k$ is the total number of photoexcited slabs.
Finally, the THz complex transmission is given by
\begin{equation}
T(\omega)\exp(i\Delta\phi(\omega))=\frac{t_{w/LD}}{t_{wo/LD}}=\frac{(Q+iP)_{wo/LD}}{(Q+iP)_{w/LD}},
\label{equ:trans}
\end{equation}
where $t_{w/LD}$ and $t_{wo/LD}$ mean the THz wave transmission with and without excitation, respectively, $Q=Re((M_{t11}+M_{t12})\sqrt{\tilde{\epsilon_{no}}}+M_{t21}+M_{t22})$ and $P=Im((M_{t11}+M_{t12})\sqrt{\tilde{\epsilon_{no}}}+M_{t21}+M_{t22})$.
The fitting of experimental $T(\omega)$ and $\Delta\phi(\omega)$ following the above mentioned procedure provides $\tilde{\sigma}(\omega)$ resulting from carriers at the photoexcited surface through
\begin{equation}
\tilde{\sigma}(\omega)=i\omega\epsilon_{0}\tilde{\epsilon}_{sur}(\omega),
\label{equ:sigma}
\end{equation}
where $\epsilon_{0}$ is the vacuum permittivity.
Note that the convergence of transmission is checked carefully by decreasing the thickness or by increasing the number of slabs.
A thickness of the photoexcited phase ($=k\times d$, $k$=100) 5 to 10 times thicker than $d_{p}$ is typically employed.

Figure 5 shows photoinduced $\tilde{\sigma}(\omega)$ of Si and YTiO$_{3}$.
$\tilde{\sigma}(\omega)$ of Si can be interpreted by the Drude model as
\begin{eqnarray}
\sigma(\omega)&=&\frac{n_{c}e\mu}{1-i\omega/\Gamma}, \label{equ:Drude}\\
\mu&=&\frac{e}{m^{*}\Gamma}, \label{equ:mob}
\end{eqnarray}
where $n_{c}$ is the carrier density, $\mu$ is the mobility, $\Gamma$ is the carrier collision rate and $m^{*}$ is the effective mass.
The photoexcitation introduces both electrons and holes; therefore, the tentatively assigned $m^{*}$ value is 0.26$m_{0}$ for electrons and 0.37$m_{0}$ for holes\cite{Exter:PRB}, where $m_{0}$ is the free-electron mass.
Hereafter, $\mu$ of each carrier is denoted as $\mu_{e}$ for electrons and $\mu_{h}$ for holes.
We have considered the following two cases, neither of which can be excluded at the present stage.
One is the two-carrier model of electrons and holes.
The other takes only electrons into consideration, assuming that holes with heavy $m^{*}$ do not contribute to $\tilde{\sigma}(\omega)$.
The solid lines in Fig.\ 5(a) represent the calculated $\tilde{\sigma}(\omega)$ for the two-carrier model and the broken lines represent the electron-only model.
The curves are in agreement with the experimental $\tilde\sigma(\omega)$.
This suggests that the itinerant carriers are certainly photogenerated in Si.
The obtained $\mu_{e}$ and $\mu_{h}$ are 2410 ($\pm$210) cm$^{2}$/Vs and 500 ($\pm$90) cm$^{2}$/Vs for the two-carrier model, and $\mu_{e}$ is 1820 ($\pm$100) cm$^{2}$/Vs for the other model.
They are roughly consistent with the literature values\cite{Sze}, but it is to be noted that $\mu_{e}$ in both models might be larger than the predicted ones.
The ambiguity of $m^{*}$ may be responsible for this deviation.

The most striking feature in $\tilde{\sigma}(\omega)$ of YTiO$_{3}$ is the negative $\sigma_{2}$.
It suggests an existence of localized carriers\cite{Turner:JPCB,Kitagawa:JPSJ,Cooke:PRB}, which is very different from Si.
The localization may arise from the on-site strong Coulomb interaction between 3d electrons in YTiO$_{3}$.
To explain $\tilde{\sigma}(\omega)$, we used the empirical Jonscher law\cite{Jonscher:Nature}, which expresses $\tilde{\sigma}(\omega)$ of many materials with hopping carriers.
The Jonscher law is given by\cite{Jonscher:Nature,Elliott:AP}
\begin{eqnarray}
\sigma_{1}(\omega)&=&\sigma_{dc}+A\omega^{s}, \label{equ:Jonscher1} \\
\sigma_{2}(\omega)&=&-A\omega^{s}\tan\frac{s\pi}{2},
\label{equ:Jonscher2}
\end{eqnarray}
where $\sigma_{dc}$ is the DC conductivity, $A$ the proportional coefficient and $s$ is restricted between 0 and 1.
As shown in Fig.\ 5(b), the solid curves from the Jonscher law seem to agree with experimental $\tilde{\sigma}(\omega)$.
In the solid curves, $s$, $\sigma_{dc}$ and $A$ are 0.95, 235 ($\pm$10) $\Omega^{-1}$cm$^{-1}$ and 2.40 ($\pm$0.05)$\times$10$^{-11}$ $\Omega^{-1}$cm$^{-1}$s$^{0.95}$, respectively.
The allowed $s$ ranges from 0.91 to 0.99, and corresponding $\sigma_{dc}$ and $A$ are 210 ($\pm$10) $\Omega^{-1}$cm$^{-1}$ and 1.41 ($\pm$0.03)$\times$10$^{-10}$ $\Omega^{-1}$cm$^{-1}$s$^{0.91}$, and 260 ($\pm$10) $\Omega^{-1}$cm$^{-1}$ and 1.43 ($\pm$0.03)$\times$10$^{-12}$ $\Omega^{-1}$cm$^{-1}$s$^{0.99}$, respectively.

Note that $\tilde{\sigma}(\omega)$ can be also fitted by a two-component model, such as the Drude-Lorentz.
The estimated photoinduced carrier number at the surface layer is about 0.015 per Ti site. 
Photoexcited YTiO$_{3}$ with the derived carrier density would be equivalent to chemically hole-doped Y$_{1-x}$Ca$_{x}$TiO$_{3}$ with $x$ much less than 0.1 given in Ref.\ 36.
The $\sigma_{1}(\omega)$ spectrum of Y$_{1-x}$Ca$_{x}$TiO$_{3}$ in this composition region is very different than a Drude response.
Therefore, it would be difficult to expect a Drude component to exist.
Clarifying this point might require broadband spectroscopic information obtained under photoexcitation or the temperature dependence of $\tilde{\sigma}(\omega)$.

Since long relaxation times, $\tau$, are observed in both samples, a diffusion or a surface-recombination process, making the carrier distribution a non-exponential decay type, must be considered, and the analysis method modified.
The carrier number $n(z)$ along the THz wave propagation in a quasi-equilibrium state is obtained using a one-dimensional diffusion equation\cite{Vaitkus:PSS} as follows:
\begin{equation}
\frac{\partial n(z,t)}{\partial t}=D\frac{\partial^{2} n(z,t)}{\partial z^{2}}-\frac{n(z,t)}{\tau}+\delta(t)\exp\left(-\frac{z}{d_{p}}\right),
\label{equ:1-diffusion}
\end{equation}
where $n(z,t)$ depends on the time $t$ and the position $z$ along the THz wave propagation, and $\delta(t)$ is the $\delta$-function.
$D$ is the diffusion coefficient and given by
\begin{equation}
D=\frac{\mu_{bi}k_{B}T_{s}}{|e|},
\label{equ:diffcoeff}
\end{equation}
where $1/\mu_{bi}$ is equal to $1/\mu_{e}+1/\mu_{h}$, $k_{B}$ is the Boltzmann constant and $T_{s}$ is the sample temperature equal to 300 K.
The solution\cite{Vaitkus:PSS} of eq. (\ref{equ:1-diffusion}) is
\begin{eqnarray}
n(z,t)&=&\exp\left(-\frac{z^{2}}{4Dt}\right)\left\{\frac{1}{2}\left[f\left(\frac{\sqrt{Dt}}{d_{p}}-\frac{z}{2\sqrt{Dt}}\right)+\frac{\frac{D}{d_{p}}+v_{s}}{\frac{D}{d_{p}}-v_{s}}f\left(\frac{\sqrt{Dt}}{d_{p}}+\frac{z}{2\sqrt{Dt}}\right)\right]\right. \nonumber \\
&&\left.-\frac{v_{s}}{\frac{D}{d_{p}}-v_{s}}f\left(v_{s}\sqrt{\frac{t}{D}}+\frac{z}{2\sqrt{Dt}}\right)\right\}\exp\left(-\frac{t}{\tau}\right),
\label{equ:diff-sol}
\end{eqnarray}
where $v_{s}$ is the surface recombination velocity and $f(z)$ is related to the error function by $f(z)=\exp(z^{2})(1-$erf$(z))$.
The carrier number in the quasi-equilibrium state requires the integration of $n(z,t)$ with respect to $t$,
\begin{equation}
n(z)=\int_0^\infty n(z,t)dt.
\label{equ:carnumber}
\end{equation}
Therefore, with the assumption of a conduction model and the knowledge of $n(z)$ determined by appropriate $\mu_{bi}$ and $v_{s}$, $T(\omega)e^{i\Delta\phi(\omega)}$ can be calculated using eq. (\ref{equ:trans}).
In this case, eq. (\ref{equ:epsilon}) is replaced by
\begin{equation}
\tilde{n}_{j}^{2}=\tilde{\epsilon}_{no}+\tilde{\epsilon}_{max}\frac{n(z_{j})}{n_{max}},
\label{equ:new-epsilon}
\end{equation}
where $n_{max}$ and $\tilde{\epsilon}_{max}$ are $n(z)$ and $\tilde{\epsilon}$ of the highest carrier-concentration layer, respectively.

After the determination of $\mu_{bi}$ of Si (= 346 cm$^{2}$/Vs) using the literature values\cite{Sze}, $v_{s}$ is varied between 1$\times$10$^{4}$ cm/s and 1$\times$10$^{6}$ cm/s.
Representative $n(z)$ normalized at $n_{max}$ are shown in the inset of Fig.\ 6(a).
The $n_{max}$ is observed around 1$\sim$2 $\mu$m.
Assuming that both electrons and holes obeying the Drude conductivity are responsible for $\tilde{\sigma}(\omega)$, $T(\omega)e^{i\Delta\phi(\omega)}$ is confirmed as being consistent with experimental data for both $n(z)$ (see Fig.\ 6(a)).
The estimated $n_{c}$ is 5.2 ($\pm$0.3)$\times$10$^{16}$ cm$^{-3}$ and is comparable to that obtained by the previous model.
This indicates that, at the highest carrier-concentration layer almost the same $n_{c}$ can be obtained, irrespective of the carrier distribution decay type.
For YTiO$_{3}$, both $v_{s}$ and $\mu_{bi}$ are unknown parameters.
The wide range sweep of $v_{s}$ and $\mu_{bi}$ gives various $n(z)$ curves as depicted in the inset of Fig.\ 6(b) with peak positions around 0.1 $\mu$m.
For each $n(z)$, the experimental $T(\omega)e^{i\Delta\phi(\omega)}$ is well reproduced by the Jonscher law, where $s$ is restricted within the same range obtained in Fig.\ 5(b) (0.91$\sim$0.99).
Typical examples are shown in Fig.\ 6(b) with $s$ of 0.95. The other parameters ($\sigma_{dc}$ in $\Omega^{-1}$cm$^{-1}$ and $A$ in $\Omega^{-1}$cm$^{-1}$s$^{0.95}$) for the dotted-solid, solid and broken lines are 125($\pm$10) and 1.40($\pm$0.05)$\times$10$^{-11}$, 125($\pm$10) and 1.2($\pm$0.1)$\times$10$^{-11}$, and 90($\pm$10) and 1.0($\pm$0.1)$\times$10$^{-11}$, respectively.
$\sigma_{1}(\omega)$ and $\sigma_{2}(\omega)$ calculated from the parameters are half to two-thirds of those in Fig.\ 5(b).
Thus, for YTiO$_{3}$, the $\tilde{\sigma}(\omega)$ extracted from the model with exponentially-decaying carrier distribution roughly represents the highest carrier-concentration layer in the model using eq. (\ref{equ:new-epsilon}).

The photoinduced carrier number at the highest carrier-concentration layer in YTiO$_{3}$ is calculated as 0.015 per Ti site\cite{carriernum}.
It can be proposed, therefore, that a phase with localized carriers would emerge initially at the photogeneration of the metallic phase in Mott insulators.
Photoexcitation creates both electrons and holes, which differs from chemical doping, and a comparison of $\tilde{\sigma}(\omega)$ between photoexcited YTiO$_{3}$ and a hole-doped Y$_{1-x}$Ca$_{x}$TiO$_{3}$ is discussed.
The absolute value of $\sigma_{1}(\omega)$ of photoexcited state might be much larger than that of corresponding Y$_{1-x}$Ca$_{x}$TiO$_{3}$ if the extrapolation of $\sigma_{1}(\omega)$ in Y$_{1-x}$Ca$_{x}$TiO$_{3}$ is carried out toward the THz energy.
The preservation of spectral weight implies that the localization energy of photoinduced carriers would be much lower than for holes in Y$_{1-x}$Ca$_{x}$TiO$_{3}$, even if both holes and electrons contribute to $\tilde{\sigma}(\omega)$ in photoexcited YTiO$_{3}$.
As it is not clear that the large difference in localization energy originates from only holes in such a low-carrier system, it is plausible that electrons with small localization energies also contribute to $\tilde{\sigma}(\omega)$.
Therefore $\tilde{\sigma}(\omega)$ of photoexcited YTiO$_{3}$ would be supported by bound electrons as well as holes.

In a halogen-bridged Ni one-dimensional chain compound [Ni(chxn)$_{2}$Br]Br$_{2}$ (chxn = cyclohexanediamine), which is compared with YTiO$_{3}$ composed of a three-dimensional Ti network, the localized $\sigma_{1}$ is determined at a lower excitation density\cite{Iwai:PRL}.
Despite being in a different energy region, carrier localization in photoexcited Mott insulators at low excitation densities may be the general phenomenon, irrespective of the dimensionality.

The fact that one-dimensional Mott insulators, such as [Ni(chxn)$_{2}$Br]Br$_{2}$\cite{Iwai:PRL} and Sr$_{2}$CuO$_{3}$\cite{Ogasawara:PRL}, exhibit $\tau$ in the order of pico-seconds may suggest that dimensionality is a decisive factor of $\tau$. 

Figure 7 shows $T(\omega)$ and $\Delta\phi(\omega)$ obtained by OPTP experiments using a femtosecond-pulse laser (1.55 eV) for YTiO$_{3}$.
Since the period of optical-pump arrival time (13 ns) is much shorter than $\tau$, the photoinduced carriers are also in quasi-equilibrium state.
In both polarizations of $E_{1.55eV}$, it is found that the degree of variation from the unexcited state in THz wave amplitude and phase is larger for $E_{THz}||b$ within the measured THz energy range.
This implies that the absolute values of $\sigma_{1}$ and $\sigma_{2}$ for $E_{THz}||b$ are larger than those for $E_{THz}||c$.
The anisotropy would reflect the crystal symmetry of YTiO$_{3}$ or the 3$d$-orbital state at Ti site.
Comparing with the OPTP results to those using the CW LD, the $\tilde{\sigma}(\omega)$ does not seem to depend strongly on the optical-photon energy.

\section{Summary}

We have optically characterized photoinduced carriers of Mott insulator YTiO$_{3}$ at low excitation densities in the THz regime by OPTP measurements, and compared the experimental results with those for band semiconductor Si.
The $\tau$ of the photoinduced carriers in YTiO$_{3}$ is about 1.5 ms.
The inhomogeneous carrier distribution along the THz wave propagation can be treated accurately using the transfer-matrix method.
This method successfully determined $\tilde{\sigma}(\omega)$ of the highest carrier-concentration layer under the quasi-equilibrium states.
YTiO$_{3}$ shows localized $\tilde\sigma(\omega)$, possibly with the Jonscher law, whereas Si exhibits the Drude response.
Anisotropic $\tilde{\sigma}(\omega)$ in YTiO$_{3}$ is determined.
Our study demonstrates that localized carriers might play an important role in the incipient formation of metallic phases in photoexcited Mott insulators.
Although the exact origin of the localization in YTiO$_{3}$ remains an open question, THz-TDS under photoexcitation with another photon energy or for another Mott insulator might provide the answer.
We note here that a preliminary THz-TDS experiment of YTiO$_{3}$ excited by a CW LD of 1.9 eV also leads to localized $\tilde{\sigma}(\omega)$.

\ack
This work was supported by Casio Science Foundation, and Strategic Information and Communications R\&D Promotion Programme of Ministry of Public Management, Home Affairs, Posts and Telecommunications.

\clearpage
\begin{figure}[hbtp]
\begin{center}
\includegraphics[width=\linewidth]{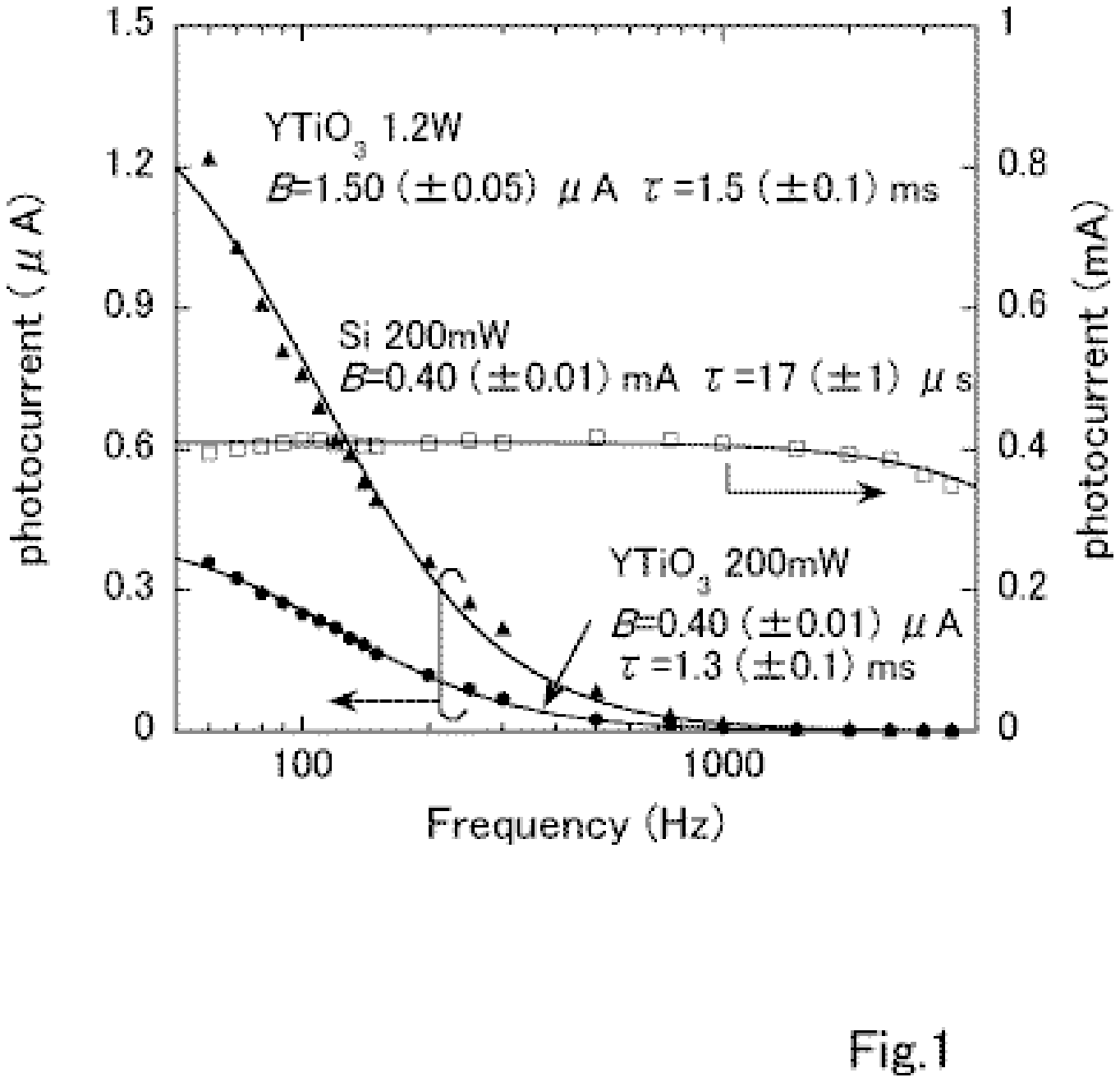}
\caption{Photocurrent as a function of modulation frequency for Si and YTiO$_{3}$. The solid lines are $B/(1+(\omega\tau)^{2})$ calculated with $B$ and $\tau$ denoted in the figure.}
\end{center}
\end{figure}

\clearpage
\begin{figure}[hbtp]
\begin{center}
\includegraphics[width=\linewidth]{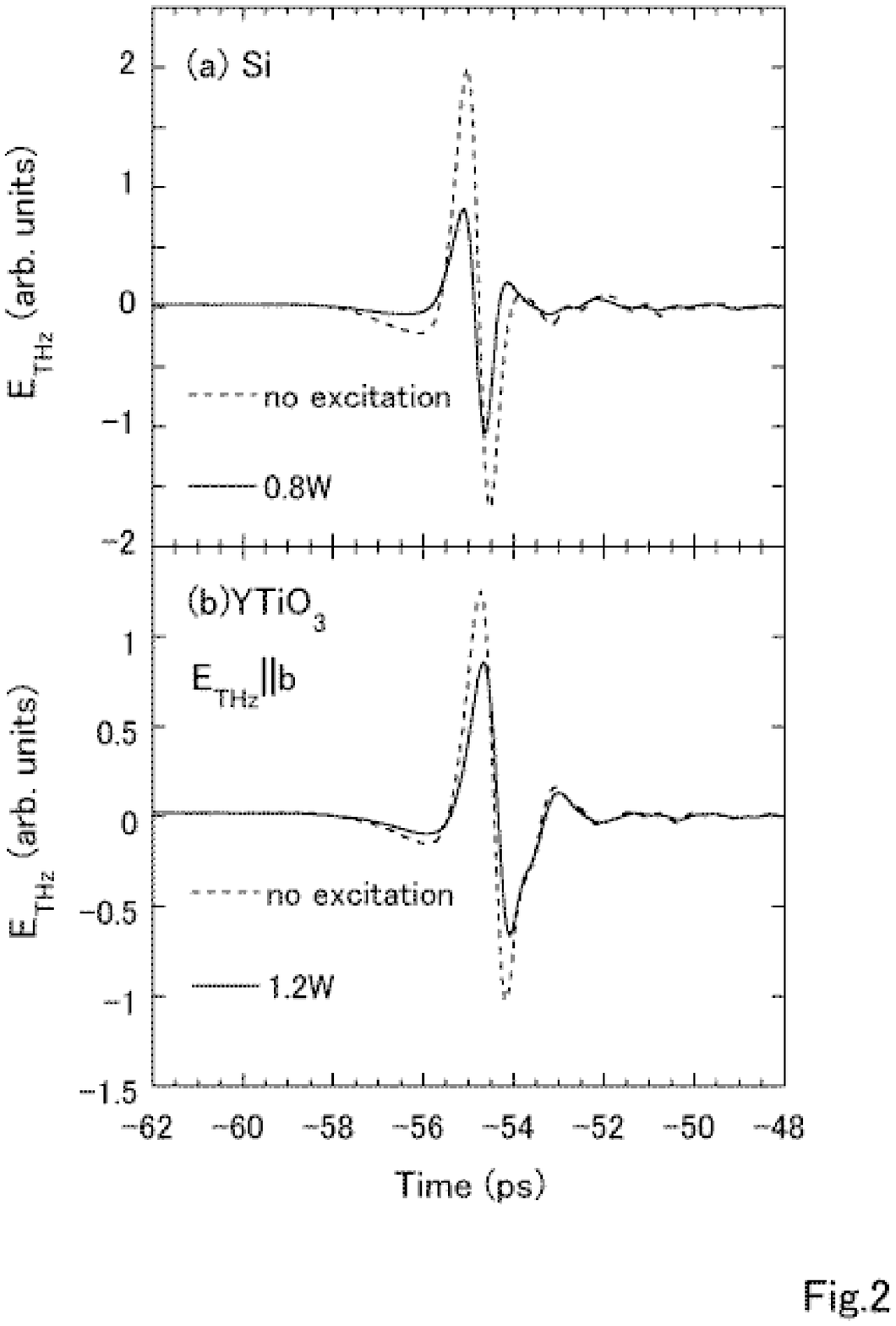}
\caption{Time evolution of $E_{THz}$ transmitted through (a) Si and (b) YTiO$_{3}$ with (solid lines) and without (broken lines) excitations (1.47 eV). The polarization of $E_{THz}$ is parallel to the $b$-axis in YTiO$_{3}$. The excitation power is 0.8 W for Si and 1.2 W for YTiO$_{3}$.}
\end{center}
\end{figure}

\clearpage
\begin{figure}[hbtp]
\begin{center}
\includegraphics[width=\linewidth]{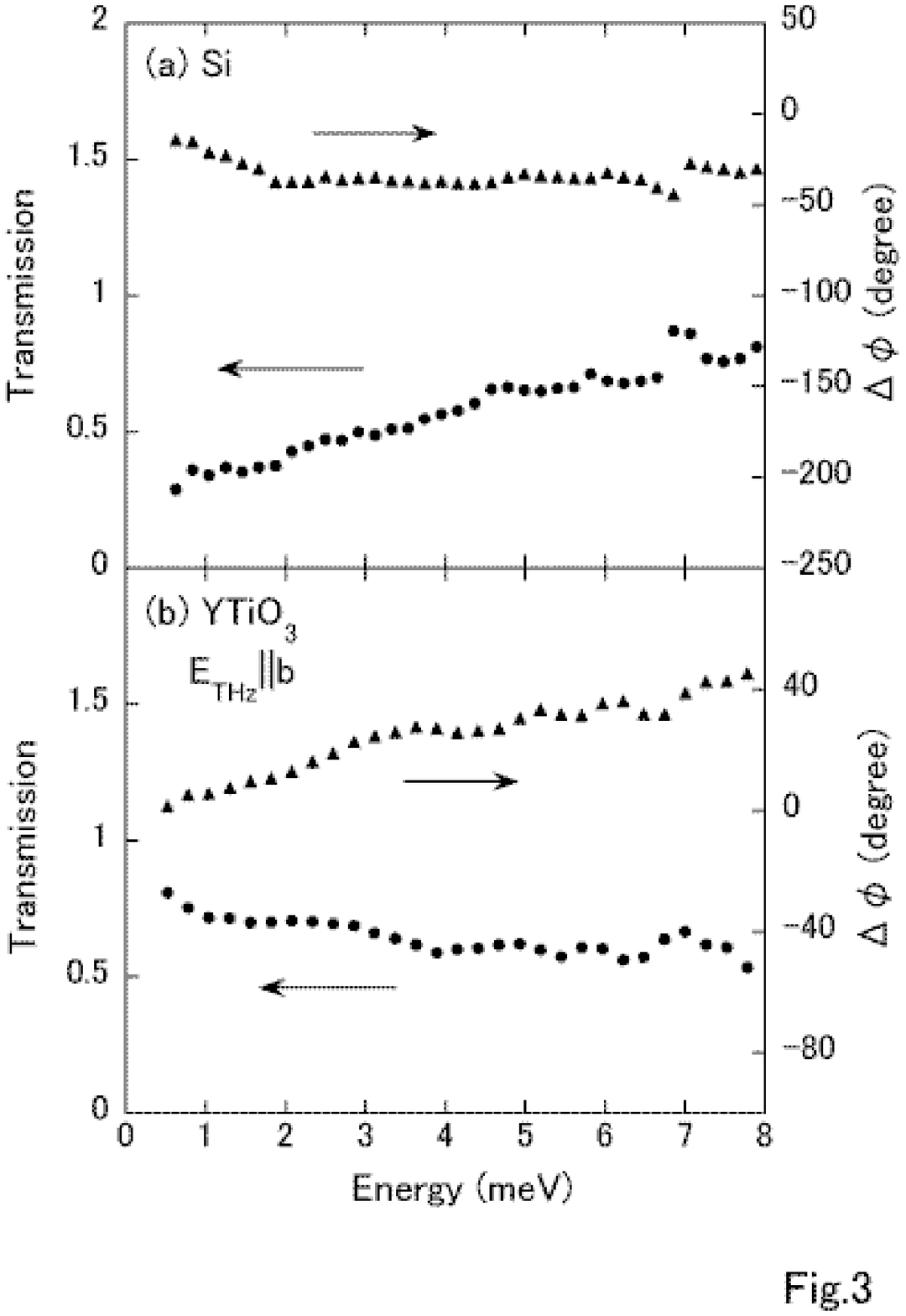}
\caption{THz energy dependence of transmission and phase shift for (a) Si and (b) YTiO$_{3}$. The polarization of $E_{THz}$ is parallel to the $b$-axis in YTiO$_{3}$.}
\end{center}
\end{figure}

\clearpage
\begin{figure}[hbtp]
\begin{center}
\includegraphics[width=0.7\linewidth]{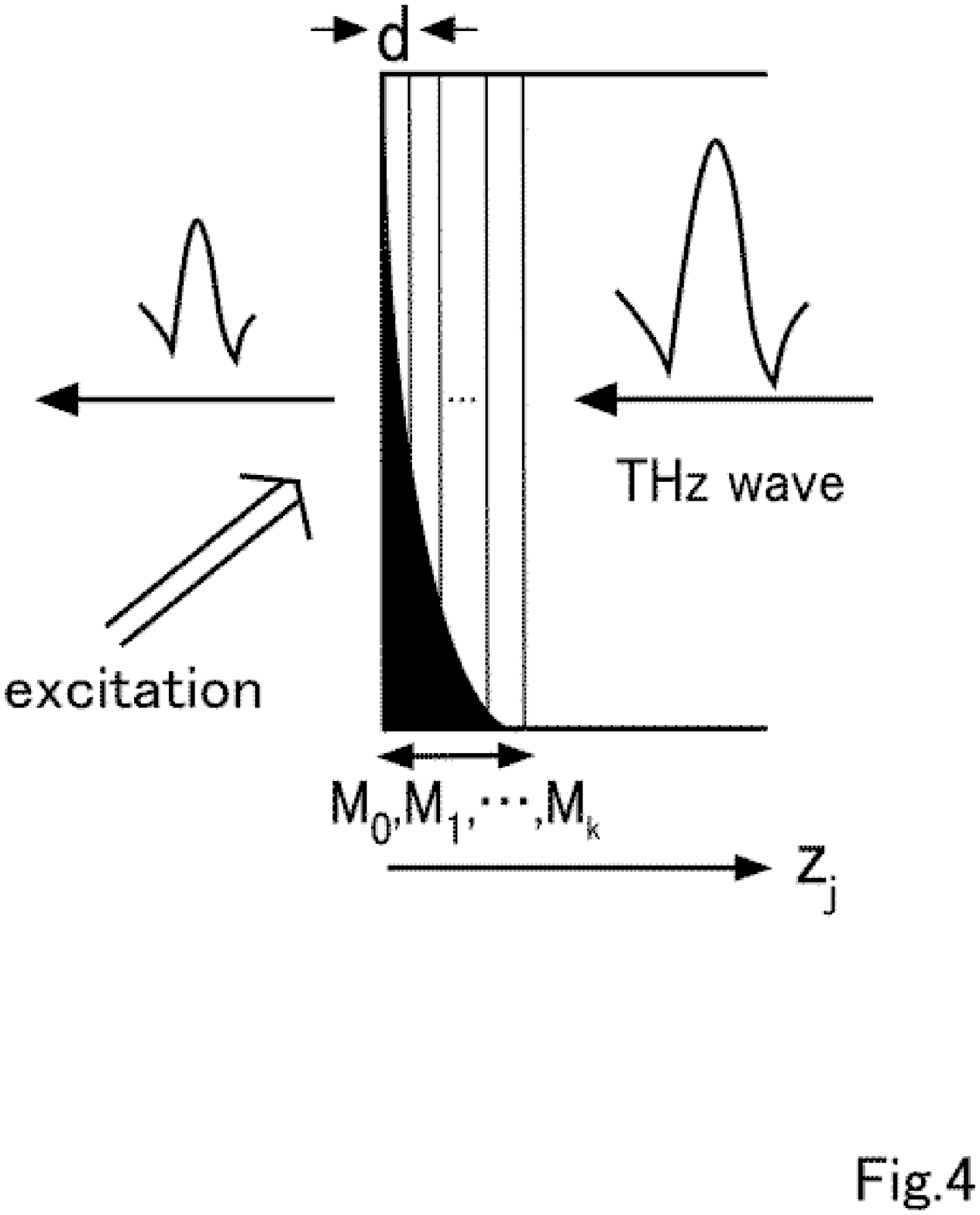}
\caption{Illustration of a photoexcited sample to explain the transfer-matrix method. The coloured area demonstrates an exponentially-decaying carrier density. The area is divided into many thin slabs with thickness $d$. The total number of thin slabs is $k$, and the transfer-matrix of each slab is expressed as $M_{j}$ ($j=0,1,...,k$). The depth from the excited surface into the sample along the THz wave propagation is denoted as $z_{j}$.}
\end{center}
\end{figure}

\clearpage
\begin{figure}[hbtp]
\begin{center}
\includegraphics[width=0.8\linewidth]{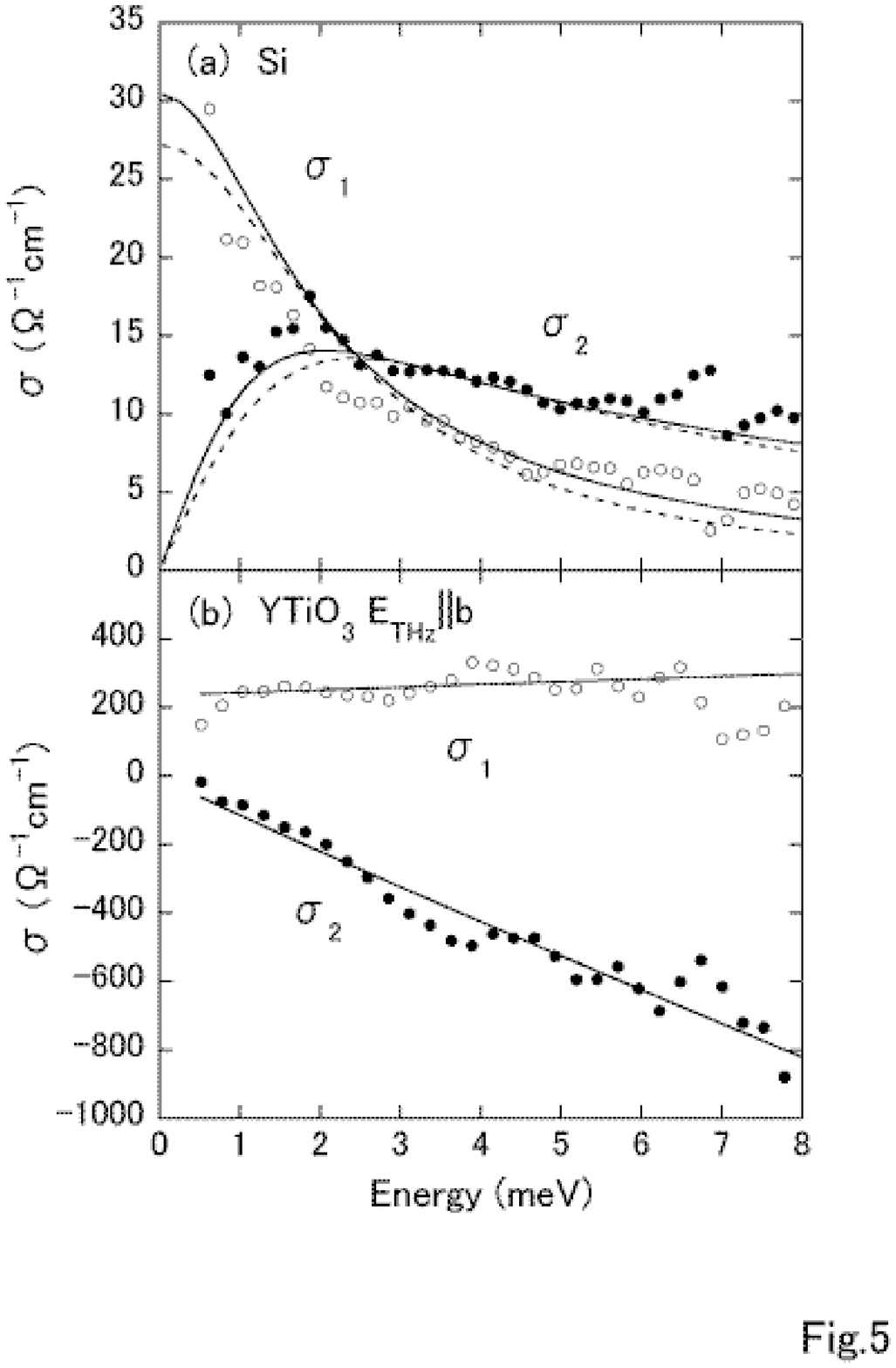}
\caption{THz complex conductivity of (a) Si and (b) YTiO$_{3}$ at the photoexcited surface. The real and the imaginary part of conductivity correspond to $\sigma_{1}$ and $\sigma_{2}$, respectively. The solid and broken curves in (a) are $\tilde{\sigma}(\omega)$ calculated by the Drude models with $n_{c}$=6.6 ($\pm$0.3)$\times$10$^{16}$ cm$^{-3}$, $\mu_{e}$=2410 ($\pm$210) cm$^{2}$/Vs and $\mu_{h}$=500($\pm$90) cm$^{2}$/Vs for the solid curves (two-carrier model of electrons and holes), and with $n_{c}$=9.3($\pm$0.4)$\times$10$^{16}$ cm$^{-3}$ and $\mu_{e}$=1820 ($\pm$100) cm$^{2}$/Vs for the broken ones (only electrons under consideration), respectively. In the two-carrier model, the same $n_{c}$ is assumed for each carrier. The solid curves in (b) are calculated $\tilde{\sigma}(\omega)$ using the Jonscher law with $\sigma_{dc}=$235 ($\pm$10) $\Omega^{-1}$cm$^{-1}$, $A=$2.40 ($\pm$0.05)$\times$10$^{-11}$ $\Omega^{-1}$cm$^{-1}$s$^{0.95}$ and $s=$0.95.}
\end{center}
\end{figure}

\clearpage
\begin{figure}[hbtp]
\begin{center}
\includegraphics[width=0.8\linewidth]{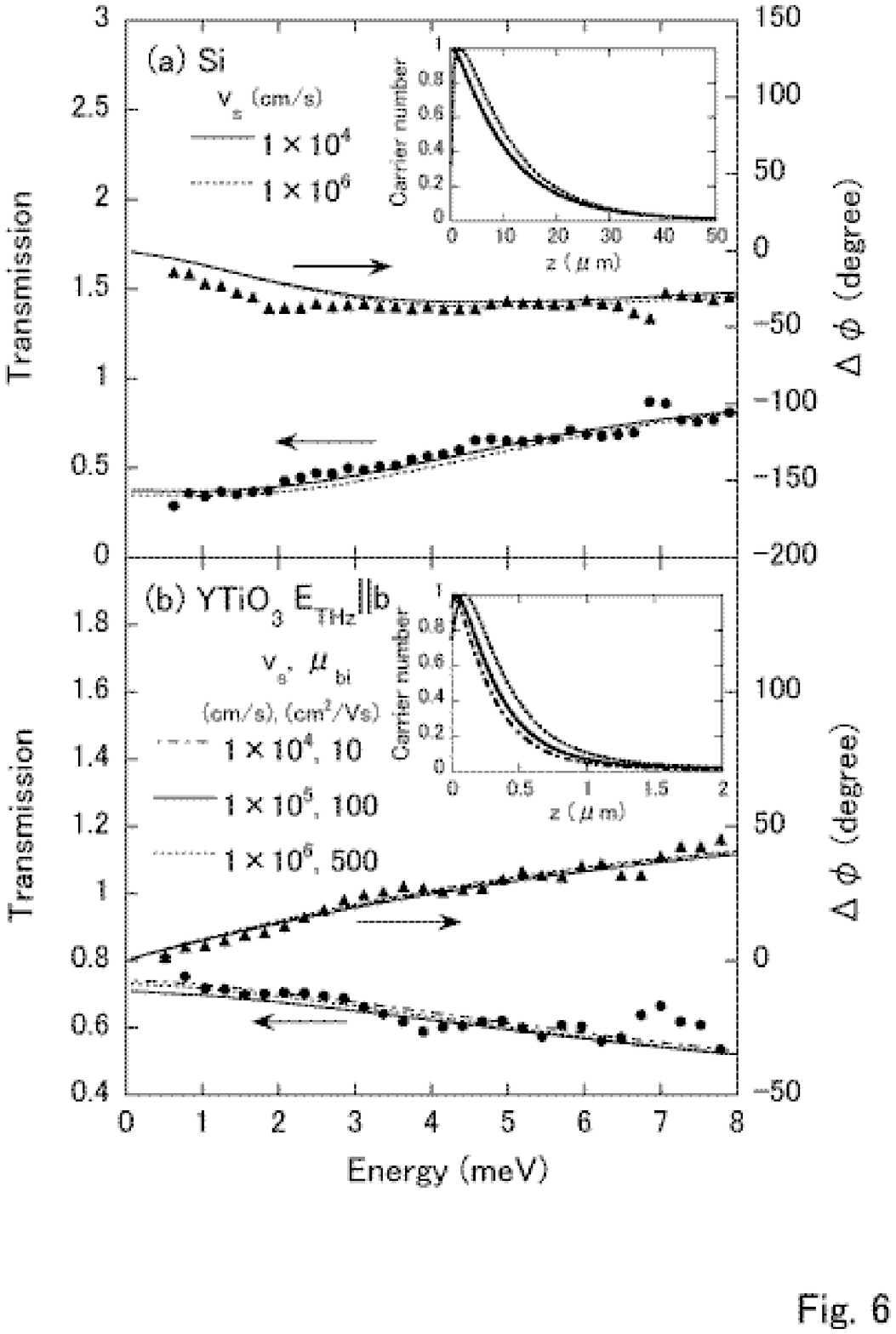}
\caption{THz energy dependence of transmission and phase shift for (a)Si and (b)YTiO$_{3}$ analysed by the model with non-exponential carrier-distribution decay. The evaluated parameters $v_{s}$ and $\mu_{bi}$ are listed in the figure. In (a), Drude conductivity characterized by $n_{c}=$5.2 ($\pm$0.3)$\times$10$^{16}$ cm$^{-3}$ and the literature values of $\mu_{e}=$1500 cm$^{2}$/Vs and $\mu_{h}=$450 cm$^{2}$/Vs is assumed. The conduction model employed in (b) is the Jonscher law ($s$=0.95), where $\sigma_{dc}$ in $\Omega^{-1}$cm$^{-1}$ and $A$ in $\Omega^{-1}$cm$^{-1}$s$^{0.95}$ are 125 ($\pm$10) and 1.40 ($\pm$0.05)$\times$10$^{-11}$, 125 ($\pm$10) and 1.2 ($\pm$0.1)$\times$10$^{-11}$, and 90 ($\pm$10) and 1.0 ($\pm$0.1)$\times$10$^{-11}$ for the dotted-solid, solid and broken curves, respectively. The insets of (a) and (b) show $z$ dependencies of the carrier numbers normalized at $n_{max}$.}
\end{center}
\end{figure}

\clearpage
\begin{figure}[hbtp]
\begin{center}
\includegraphics[width=\linewidth]{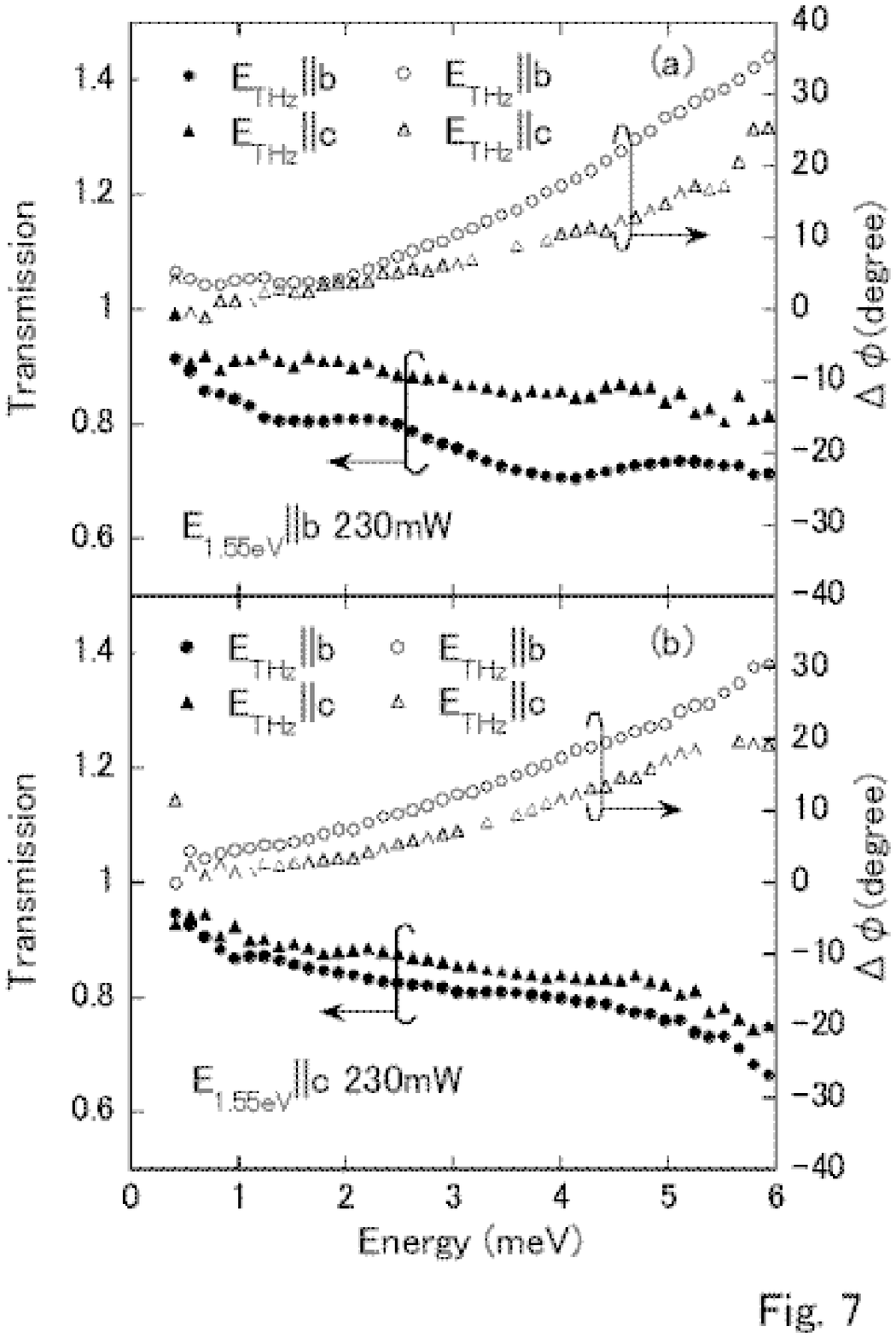}
\caption{THz energy dependence of transmission and phase shift of YTiO$_{3}$ obtained by OPTP experiments, using the femtosecond-pulse laser (1.55 eV) under $E_{THz}||$b and $E_{THz}||$c for (a)$E_{1.55eV}||$b and (b)$E_{1.55eV}||$c. The power of the optical-pump pulse is 230 mW.}
\end{center}
\end{figure}


\begin{thebibliography}{99}
\bibitem{thermal}
The temperature rise $\Delta T_{s}$ under an equilibrium condition is calculated as follows. If all of the excitation power $P$ is transformed into heat in a sample, $\Delta T_{s}$ is related to $P$ by $\Delta T_{s}=PL/S\kappa$ where $L$ is the sample length, $S$ is the illuminated area\cite{carriernum} of 0.5 cm$^{2}$ and $\kappa$ is the thermal conductivity of the sample (150 W/mK for Si\cite{Sze}). With the lack of $\kappa$ data for YTiO$_{3}$, we employ underestimated $\kappa=$10 W/mK.
\bibitem{carriernum}
The photoinduced carrier density $n_{c}$ at $P$ is tentatively expressed by $n_{c}=\frac{1}{1-e^{-\hbar\omega/P\tau}}\times F\times\frac{1}{S\cdot d_{eff}}$ where the first term indicates the carrier accumulation resulting from longer $\tau$ at photon energy $\hbar\omega$ of 1.47 eV, the second one is the Fresnel loss ($\sim$ 0.8 for YTiO$_{3}$ and $\sim$ 0.7 for Si). $d_{eff}$ means the effective $d_{p}$ defined by $\int_0^\infty n(z)dz$/$n_{max}$ and is equal to 13 $\mu$m for Si and 0.55 $\mu$m for YTiO$_{3}$. This equation is initially applied to the experimental result of Si (Fig. 6(a)) with $n_{c}=$5.2$\times$10$^{16}$ cm$^{-3}$, in which $S$ is determined to be 0.5 cm$^{2}$. Using the $S$, the carrier number of YTiO$_{3}$ can be derived.
\end{thebibliography}
\end{document}